\documentclass[epj]{svjour}

\usepackage{hyperref}
\usepackage{doi}
\usepackage{graphics}
\usepackage{epsfig}
\usepackage{soul}
\usepackage{color}
\usepackage[utf8]{inputenc}

\usepackage{mathalfa}

\usepackage[sort&compress,numbers]{natbib}

\begin{document}
\title{Heavy hadron molecules in effective field theory:\\
  the emergence of exotic nuclear landscapes}
\author{Manuel Pavon Valderrama
}                     
\offprints{}          
\institute{School of Physics and Nuclear Energy Engineering, \\
International Research Center for Nuclei and Particles in the Cosmos and \\
Beijing Key Laboratory of Advanced Nuclear Materials and Physics, \\ 
Beihang University, Beijing 100191, China, \email{mpavon@buaa.edu.cn}
}
\date{Received: date / Revised version: date}
%
\abstract{
  Heavy hadron molecules were first theorized from a crude analogy with
  the deuteron and the nuclear forces binding it, a conjecture which
  was proven to be on the right track after the discovery of the $X(3872)$.
  However, this analogy with nuclear physics has not been seriously
  exploited beyond a few calculations in the two- and three-body sectors,
  leaving a great number of possible theoretical consequences unexplored.
  Here we show that nuclear and heavy hadron effective field theories are
  formally identical: using a suitable notation, there is no formal
  difference between these two effective field theories.
  For this, instead of using the standard heavy superfield notation,
  we have written the heavy hadron interactions directly
  in terms of the light quark degrees of freedom.
  We give a few examples of how to exploit this analogy,
  e.g. the calculation of the two-pion exchange diagrams.
  Yet the most relevant application of the present idea is the conjecture
  of exotic nuclear landscapes, i.e. the possibility of few heavy hadron
  bound states with characteristics similar
  to those of the standard nuclei.
\PACS{
      {PACS-key}{describing text of that key}   \and
      {PACS-key}{describing text of that key}
     } 
} 
\titlerunning{the emergence of exotic nuclear landscapes}

\maketitle
\section{Introduction}
\label{sec:intro}


Heavy hadron molecules, i.e. bound states of heavy hadrons, were first
conjectured from a direct analogy to the deuteron
in nuclear physics~\cite{Voloshin:1976ap}.
From a phenomenological perspective the exchange of light mesons
generates a force that can bind not only nucleons, but also other
heavy hadrons together (at least if they contain a light quark).
The experimental discovery of the $X(3872)$
by the Belle collaboration~\cite{Choi:2003ue}
put this conjecture in the forefront of hadron physics.
Among the different explanations of the $X(3872)$, the most solid one is
that it is a $D^* \bar{D}$ bound state~\cite{Tornqvist:2003na,Voloshin:2003nt,Braaten:2003he}.
Circumstantial evidence is provided by the closeness of the $X$ to the
$D^{*0} D^0$ threshold, while the isospin breaking decays
into $J/\Psi\,2 \pi$ and $J/\Psi\,3 \pi$ provide a much stronger case
for the molecular nature of the $X$~\cite{Gamermann:2009fv,Gamermann:2009uq,Hanhart:2011tn}.
Yet the most compelling evidence would be the $D^{0} D^{0} \pi$ and
$D^{0} D^{0} \gamma$ decays~\cite{Voloshin:2005rt,Fleming:2007rp,Guo:2014hqa,Guo:2019qcn}, which have not been
experimentally measured yet.
The subsequent experimental discovery of a few additional molecular
candidates, such as the $Z_b$'s~\cite{Belle:2011aa,Garmash:2014dhx},
the $Z_c$'s~\cite{Ablikim:2013mio, Liu:2013dau,Ablikim:2013wzq, Ablikim:2014dxl}
and the $P_c$'s~\cite{Aaij:2015tga,Aaij:2019vzc}
(theorized to be $B^*\bar{B}$/$B^*\bar{B}^*$~\cite{Voloshin:2011qa,Cleven:2011gp}, $D^*\bar{D}$/$D^*\bar{D}^*$~\cite{Wang:2013cya,Guo:2013sya} 
and $\bar{D} \Sigma_c$ / $\bar{D}^* \Sigma_c$ / $\bar{D}^* \Sigma_c^*$~\cite{Roca:2015dva,He:2015cea,Xiao:2015fia,Chen:2015loa,Chen:2015moa,Chen:2019bip,Chen:2019asm,Liu:2018zzu,Liu:2019tjn,Xiao:2019aya} /
$\bar{D} \Lambda_{c}(2595)$~\cite{Burns:2015dwa,Geng:2017hxc} /
$[{c\bar c}]\,p$~\cite{Meissner:2015mza,Eides:2019tgv} molecules, respectively),
have further increased the interest
in heavy hadron bound states.

Heavy hadron molecules are probably among the most studied types of
exotic hadrons, i.e. hadrons that do not fit into
the quark-antiquark or three-quark picture.
This is not surprising if we consider the fertility and flexibility of
the molecular hypothesis, see Ref.~\cite{Guo:2017jvc}
for a recent review.
Among the achievements of the molecular hypothesis we can count
the prediction of the $X(3872)$ by T\"ornqvist~\cite{Tornqvist:1993ng}
and of the hidden charm pentaquarks~\cite{Wu:2010jy,Wu:2010vk,Wu:2010rv,Xiao:2013yca,Wang:2011rga,Yang:2011wz,Karliner:2015ina}.
Nonetheless there are important gaps, of which the most evident is a general
lack of theoretical coherence in the molecular picture (particularly
if we compare it with quarkonium studies~\cite{Eichten:1978tg,Eichten:1979ms,Brambilla:1999xf,Brambilla:2004jw,Brambilla:2010cs}),
which often manifests as applications of the molecular hypothesis
in a case-by-case basis that relies on ad-hoc methods.
This points towards the necessity of increasing systematicity, which includes
the determination of the plausible molecular spectrum (within uncertainties)
and working out the observable differences between a molecular
and a compact hadron.
In this regard the application of effective field theory (EFT) ideas to
the theoretical exploration of molecular states has been
indeed a welcomed addition.

The study of hadronic molecules began as an offspring of nuclear physics.
The most systematic attempts to understand them have been indeed based
on nuclear physics, including pionless EFT (nuclear physics~\cite{vanKolck:1998bw,Chen:1999tn} / hadronic molecules~\cite{Braaten:2003he,Mehen:2011yh,Guo:2013sya,Liu:2018zzu,Liu:2019tjn}),
pionful EFT (nuclear physics~\cite{Weinberg:1990rz,Weinberg:1991um,Kaplan:1998tg,Kaplan:1998we,Nogga:2005hy,Birse:2005um,Valderrama:2009ei,Valderrama:2011mv,Long:2011qx,Long:2011xw,Long:2012ve} / hadronic molecules~\cite{Fleming:2007rp,Baru:2011rs,Valderrama:2012jv,Nieves:2012tt,Baru:2015nea,Baru:2016iwj,Baru:2017gwo,Meng:2019ilv}),
and the one boson exchange (OBE) model (nuclear physics~\cite{Machleidt:1987hj,Machleidt:1989tm} / hadronic molecules~\cite{Liu:2008tn,Sun:2011uh,Chen:2015loa,Liu:2018bkx,Wang:2019nwt}).
At this point we have used the word systematic in a broad sense,
meaning a coherent or unified description instead of
the more specific meaning it has within
the EFT formalism.
The ideas developed in nuclear physics can still contribute to the exploration
of these interesting objects.
In the present manuscript we want to explore this analogy further and extend it,
which as we will see will bring us to a few useful calculations and
new predictions.

The point of the present manuscript is that the relation between the nuclear
and heavy hadron sectors is not simply an analogy, but rather a formal
equivalence for heavy hadrons containing one S-wave light quark
(for instance, the $D$, $D^*$ charmed mesons and the
$\Xi_{cc}$, $\Xi_{cc}^*$ doubly charmed baryons).
Conversely, heavy hadrons that contain other more complex light quark
configurations can be interpreted as a formal extension of nuclear physics.
This equivalence is particularly clear within a suitable notation
for the heavy hadron fields inspired in the quark model
that has been used in the past, e.g. in Ref.~\cite{Manohar:1992nd},
and that we recover here.
This in turn makes it easy to translate a few results of
nuclear EFT to heavy hadrons.
In particular and for illustrative purposes, we will derive the leading
two-pion exchange potential for the heavy mesons
and doubly heavy baryons. 

Yet the formal equivalence we will show begs a more far-reaching question:
is there an equivalent of nuclear physics in the exotic sector?
That is, do doubly heavy baryons form bound states similar to nuclei?
We speculate with the possibility of a charming nuclear landscape, where
instead of nucleons we have doubly charmed baryons.
But phenomenological arguments suggest that the conditions for doubly charmed
baryon systems are not conducive to the existence of this particular
exotic nuclear landscape.
While the existence of $A = 2,3,4$ bound states of $A$ doubly charmed baryons
might be possible, for $A \geq 5$ this seems highly unlikely.
However this does not preclude the possibility of exotic nuclear landscapes
composed of charmed-bottom and doubly bottom baryons, though these systems
will not be considered in detail in the present manuscript.

The manuscript is structured as follows: in Sect.~\ref{sec:light-quark}
we present the light-quark notation we advocate in this manuscript and
compare it with the standard heavy superfield notation.
In Sect.~\ref{sec:equivalence} we explain the formal equivalence
between the S-wave heavy mesons / doubly heavy baryons
and the nucleons.
In Sect.~\ref{sec:pions} we exploit the previous equivalence to derive
the one- and leading two-pion exchange potentials for the aforementioned
heavy hadron systems without a single calculation.
In Sect.~\ref{sec:exotic} we conjecture the existence of exotic nuclear
landscapes on the basis of the previous analogy.
Finally in Sect.~\ref{sec:summary} we present our conclusions.

\section{From heavy superfields to light subfields}
\label{sec:light-quark}

Heavy quark spin symmetry (HQSS) derives from the observation that
the chromomagnetic interaction of a heavy quark $Q$
is strongly suppressed~\cite{Isgur:1989vq,Isgur:1989ed}.
The reason is that the heavy quark mass $m_Q$ is much larger
than the QCD scale $\Lambda_{\rm QCD} \sim 200-300\,{\rm MeV}$.
As a consequence the low energy interactions between heavy quarks are
independent of the spin of the heavy quarks.

This still holds true when the heavy quarks are inside hadrons:
in this case the spin dependence of the heavy-hadron interactions
comes from the light quark degrees of freedom.
This is best taken into account with a suitable notation for heavy baryons,
i.e. baryons containing a heavy quark.
The standard method is to define a superfield, as we will review in a few lines.
Yet this is not the only possible option: here we will use a notation
inspired on the quark model, in which we only explicitly write down
the light-quark degrees of freedom within the heavy hadron.

\subsection{Heavy quark spin symmetry and the heavy superfields}
\label{subsec:superfields}

The quantum numbers of a heavy hadron ($Q \bar q$, $QQ q$, $Q qq$)
stem from the coupling between the heavy and light spin degrees of
freedom within it
\begin{eqnarray}
  | J m \rangle = \sum_{m_H, m_L}\,|S_H m_H \rangle | S_L m_L \rangle \,
  \langle S_H m_H S_L m_L | J m \rangle \, , \label{eq:heavy-JM}
\end{eqnarray}
where $|S_H m_H \rangle$ and $| S_L m_L \rangle$ are the spin wave functions of
the heavy and light degrees of freedom, respectively, and 
$\langle S_H m_H S_L m_L | J m \rangle$
are the Clebsch-Gordan coefficient.
This gives rise to a family of heavy hadrons with
\begin{eqnarray}
J = |S_H - S_L|, \dots, |S_H + S_L|,
\end{eqnarray}
which according to HQSS should all have the same mass and properties.
That is, these hadrons have the same light-quark wave function,
sometimes informally referred to as the ``brown muck'',
a term introduced by Isgur~\cite{Isgur:1991xa,Flynn:1992fm}.

Alternatively, HQSS implies that the interactions of this family of
heavy hadrons are invariant under rotations of
the heavy quark spin $S_H$.
It happens that this type of rotation mixes the heavy hadrons
with different $J$ that we have defined in Eq.~(\ref{eq:heavy-JM}).
This makes it particularly difficult to write down interactions
that respect HQSS in a notation where each of the heavy hadrons
that arise from the $S_H \otimes S_L$ coupling
are written as different fields.
A solution is to group the $S_H \otimes S_L$ family of heavy hadron fields
into a unique heavy hadron superfield.

The idea of the heavy superfield notation is to group all the possible
couplings of $S_H$ and $S_L$ into a single matrix $\mathcal{M}$
--- the heavy superfield ---
with $(2 S_H + 1) \times (2 S_L + 1)$ components,
where the components of this matrix are
\begin{eqnarray}
  \mathcal{M}_{m_H, m_L} =
    | S_H m_H \rangle  | S_L m_L \rangle \, .
\end{eqnarray}
Under heavy quark spin rotations the matrix $\mathcal{M}$ transforms
exactly as the heavy quark field $Q$
\begin{eqnarray}
  Q \to e^{i \vec{S}_H \cdot \theta} Q \quad \Rightarrow \quad
  \mathcal{M} \to e^{i \vec{S}_H \cdot \theta} \mathcal{M} \, .
\end{eqnarray}
This in turn makes it easy to write interactions that respect HQSS for
the heavy hadrons conforming a particular heavy multiplet.

To be more concrete we will consider the example of the S-wave heavy mesons.
The quark content of the S-wave heavy mesons is $Q \bar{q}$,
with $Q$ a heavy quark and $\bar{q}$ a light antiquark.
Their total angular momentum is $J = 0, 1$, where the generic notation
$P$ and $P^*$ is often used to denote the $J=0$ and $J=1$ heavy meson,
respectively.
Depending on whether the heavy quark content is $Q = c, b$
we have $P = D, \bar{B}$ and $P^* = D^*, \bar{B}^*$.
As previously said, HQSS implies that the $P$ and $P^*$ heavy mesons
are degenerate and form a multiplet, where the standard methodology
to take this symmetry into account is to define the superfield
\begin{eqnarray}
  H_Q = \frac{1}{\sqrt{2}}\left[ P +
    \vec{\sigma} \cdot \vec{P}^* \right] \, ,
\end{eqnarray}
which is a $2 \times 2$ matrix (the $P$ field is implicitly multiplied by the
$2 \times 2$ identity matrix) that has good properties with respect
to heavy quark rotations, i.e.
\begin{eqnarray}
  Q \to e^{i \vec{S}_H \cdot \vec{\theta}} Q \quad \Rightarrow \quad
    H_Q \to e^{i \vec{S}_H \cdot \vec{\theta}} H_Q \, .
\end{eqnarray}
Now that the superfield $H_Q$ is defined, we can write heavy meson interactions.
For instance, if we ignore the isospin quantum numbers for simplicity,
the most general Lagrangian for contact-range interactions
(i.e. four heavy meson field vertices) without derivatives
will be
\begin{eqnarray}
  \mathcal{L} &=&
  C_a\,{\rm Tr}[H_Q^{\dagger} H_Q]\,{\rm Tr}[H_Q^{\dagger} H_Q] \nonumber \\ &+&
  C_b\,{\rm Tr}[H_Q^{\dagger} {\sigma}_i H_Q]\,\cdot\,
  {\rm Tr}[H_Q^{\dagger} {\sigma}_i H_Q] \, ,
  \label{eq:L-heavy-mesons}
\end{eqnarray}
which gives rise to a well-known contact-range potential~\cite{Mehen:2011yh,Valderrama:2012jv,Nieves:2012tt}
that explains a few regularities in the molecular spectrum~\footnote{
In particular this potential explains why the $Z_c$ and $Z_c'$ resonances appear
in pairs~\cite{Guo:2013sya}, why the same happens for the $Z_b$ and $Z_b'$
resonances~\cite{Voloshin:2011qa,Bondar:2011ev,Mehen:2011yh}, and
why we have a hidden-charm and hidden-bottom
version of them~\cite{Guo:2013sya}.
Besides,
this potential also leads to the prediction of a $2^{++}$ $D^* \bar{D}^*$
molecular partner of the $X(3872)$~\cite{Valderrama:2012jv,Nieves:2012tt},
though this partner has not been experimentally observed.
Recently a similar HQSS contact-range potential has been derived
for the $P \Sigma_Q$, $P^* \Sigma_Q$, $P \Sigma_Q^*$ and $P^* \Sigma_Q^*$
molecules~\cite{Liu:2018zzu}, which in turn has been proven to be
surprisingly useful to explain the recently observed LHCb pentaquark
trio~\cite{Aaij:2019vzc} as hadronic molecules
belonging to the same HQSS multiplet~\cite{Liu:2019tjn}.
}.
Ignoring isospin, Bose-Einstein symmetry if we are dealing with a
heavy meson-meson system and C-parity if we are dealing with a
heavy meson-antimeson system, the contact potential potential
can be consulted in Table \ref{tab:1}
for each of the six possible S-wave configurations.

\begin{table}
  \caption{The contact-range potential for the heavy meson-(anti)meson
    as constrained from HQSS.
    For simplicity only the light-spin structure of the potential is shown,
    while other factors are ignored (these factors include isospin,
    Bose-Einstein statistics in the heavy meson-meson case
    and C-parity in the heavy meson-antimeson case).
    The form of the potential is the same in the heavy meson-meson and
    heavy meson-antimeson systems, provided we use the following
    C-parity convention for the antimesons:
    $C | P \rangle = | \bar{P} \rangle$ and
    $C | P^* \rangle = | \bar{P}^* \rangle$,
    where $C$ is the C-parity operator
    (though the couplings will be different).
  }
\label{tab:1}       
\begin{center}
\begin{tabular}{|c|c|c|}
  \hline
Molecule & J & V  \\
\hline
$PP$ & 0 & $C_a$ \\
$\frac{1}{\sqrt{2}}\left( P^* P + P P^* \right)$ & 1 & $C_a + C_b$ \\
$\frac{1}{\sqrt{2}}\left( P^* P - P P^* \right)$ & 1 & $C_a - C_b$ \\
$P^* P^*$ & 0 & $C_a - 2 C_b$ \\
$P^* P^*$ & 1 & $C_a - C_b$ \\
$P^* P^*$ & 2 & $C_a + C_b$ \\
\hline
\end{tabular}
\end{center}
\end{table}

\begin{table}
  \caption{The contact-range potential for the doubly heavy
    baryon-(antibaryon) systems, as constrained from HQSS.
    We only show the light-spin structure of the potential, while ignoring
    isospin, Dirac-Fermi statistics in the doubly heavy baryon-baryon case
    and C-parity in the heavy baryon-antibaryon case.
    If HADS is taken into account, the couplings $C_a$ and $C_b$
    will be the same as in Table \ref{tab:1}.
  }
\label{tab:2}       
\begin{center}
\begin{tabular}{|c|c|c|}
  \hline 
  Molecule & $J$ & $V$ \\
\hline
$\Xi_{QQ} \Xi_{QQ}$ & 0 & $C_a - \frac{1}{3} C_b$ \\
$\Xi_{QQ} \Xi_{QQ}$ & 1 & $C_a + \frac{1}{9} C_b$ \\
\hline
$\frac{1}{\sqrt{2}}\left(
\Xi_{QQ}^* \Xi_{QQ} - \Xi_{QQ} \Xi_{QQ}^*
\right)$
& 1 & $C_a + \frac{1}{9} C_b$ \\
$\frac{1}{\sqrt{2}}\left(
\Xi_{QQ}^* \Xi_{QQ} + \Xi_{QQ} \Xi_{QQ}^*
\right)$
& 1 & $C_a +  C_b$ \\
$\frac{1}{\sqrt{2}}\left(
\Xi_{QQ}^* \Xi_{QQ} - \Xi_{QQ} \Xi_{QQ}^*
\right)$
& 2 & $C_a - \frac{5}{3} C_b$ \\
$\frac{1}{\sqrt{2}}\left(
\Xi_{QQ}^* \Xi_{QQ} + \Xi_{QQ} \Xi_{QQ}^*
\right)$
& 2 & $C_a + C_b$ \\
\hline
$\Xi_{QQ}^* \Xi_{QQ}^*$ & 0 & $C_a - \frac{5}{3} C_b$ \\
$\Xi_{QQ}^* \Xi_{QQ}^*$ & 1 & $C_a - \frac{11}{9} C_b$ \\
$\Xi_{QQ}^* \Xi_{QQ}^*$ & 2 & $C_a - \frac{1}{3} C_b$ \\
$\Xi_{QQ}^* \Xi_{QQ}^*$ & 3 & $C_a + C_b$ \\
\hline
\end{tabular}
\end{center}
\end{table}

\begin{table}
  \caption{The contact-range potential for the doubly heavy
    baryon - heavy (anti)meson systems, as constrained from HQSS.
    For simplicity only the light-spin structure of the potential is shown,
    while other factors (isospin and the C-parity convention
    for the heavy hadrons) are ignored .
    If HADS is taken into account, the couplings $C_a$ and $C_b$
    will be the same as in Tables \ref{tab:1} and \ref{tab:2}.
  }
\label{tab:3}       
\begin{center}
\begin{tabular}{|c|c|c|}
  \hline 
  Molecule & $J$ & $V$ \\
\hline
$\Xi_{QQ} {\bar P}$ & $\frac{1}{2}$ & $C_a$ \\
\hline
$\Xi_{QQ}^* {\bar P}$ & $\frac{1}{2}$ & $C_a$ \\
\hline
$\Xi_{QQ} {\bar P}^*$ & $\frac{1}{2}$ & $C_a + \frac{2}{3} C_b$ \\
$\Xi_{QQ} {\bar P}^*$ & $\frac{3}{2}$ & $C_a - \frac{1}{3} C_b$ \\
\hline
$\Xi_{QQ}^* {\bar P}^*$ & $\frac{1}{2}$ & $C_a - \frac{5}{3} C_b$ \\
$\Xi_{QQ}^* {\bar P}^*$ & $\frac{3}{2}$ & $C_a - \frac{2}{3} C_b$ \\
$\Xi_{QQ}^* {\bar P}^*$ & $\frac{5}{2}$ & $C_a + C_b$ \\
\hline
\end{tabular}
\end{center}
\end{table}

For doubly heavy baryons~\footnote{We notice in passing that as for now
  the only doubly heavy baryon that has been experimentally observed
  is the $\Xi_{cc}^{++}$ by the LHCb~\cite{Aaij:2017ueg}.}
the definition of the superfield is
\begin{eqnarray}
  T_{QQ} = \frac{1}{\sqrt{3}} \vec{\sigma} \Xi_{QQ} + \vec{\Xi}_{QQ}^* \, ,
\end{eqnarray}
with $\Xi_{QQ}$ and $\Xi_{QQ}^*$ referring to the spin-$\frac{1}{2}$ and
spin-$\frac{3}{2}$ doubly heavy baryons,
where $\sigma \cdot \Xi_{QQ}^* = 0$.
For this case the most general contact-range Lagrangian without derivatives is
\begin{eqnarray}
  \mathcal{L} &=&
  C_a\,(\vec{T}^{\dagger}_{QQ} \cdot \vec{T}_{QQ})\,
  (\vec{T}^{\dagger}_{QQ} \cdot \vec{T}_{QQ}) \nonumber \\
  &+& C_b\,(\vec{T}^{\dagger}_{QQ} \cdot \sigma_i \vec{T}_{QQ})\,
  (\vec{T}^{\dagger}_{QQ} \cdot  \sigma_i \vec{T}_{QQ}) \, ,
  \label{eq:L-doubly-heavy-baryons}
\end{eqnarray}
which has the same general structure as the one for heavy mesons.
The contact-range potential can also be read in Table \ref{tab:2} for the
ten possible S-wave configurations of two doubly heavy baryons.

Now these are related by heavy antiquark-diquark symmetry (HADS), which
in principle implies the definition of a more general superfield that
groups the $H_{\bar Q}$ and $T_{QQ}$ superfields (where the subscript
$\bar{Q}$ indicates we are considering the heavy antimeson
field)~\cite{Hu:2005gf}.
In practice it means that we can make the following substitutions
\begin{eqnarray}
  {\rm Tr}[{H}_{\bar Q}^{\dagger} {H}_{\bar Q}] &\leftrightarrow&
  (\vec{T}^{\dagger}_{QQ} \cdot \vec{T}_{QQ}) \, , \\
  {\rm Tr}[H_{\bar Q}^{\dagger} \sigma_i H_{\bar Q}] &\leftrightarrow&
  (\vec{T}^{\dagger}_{QQ} \cdot \sigma_i \vec{T}_{QQ}) \, , 
\end{eqnarray}
from which we deduce that the couplings in the contact Lagrangians of
Eqs.~(\ref{eq:L-heavy-mesons}) and (\ref{eq:L-doubly-heavy-baryons})
are identical.
Besides, from the previous substitutions we can also deduce the contact-range
Lagrangian for the interaction of heavy mesons and doubly charmed baryons
\begin{eqnarray}
  \mathcal{L} &=&
  C_a\,{\rm Tr}[H_{\bar Q}^{\dagger} H_{\bar Q}]\,
  (\vec{T}^{\dagger}_{QQ} \cdot \vec{T}_{QQ})
  \nonumber \\
  &+& C_b\,{\rm Tr}[H_{\bar Q}^{\dagger} \sigma_i H_{\bar Q}]\,
  (\vec{T}^{\dagger}_{QQ} \cdot  \sigma_i \vec{T}_{QQ}) \, ,
\end{eqnarray}
which leads to the contact-range potential of Table \ref{tab:3}.
This Lagrangian has in turn been used in the past for predicting
the existence of triply-charmed molecular pentaquarks
from the assumption that the $X(3872)$ is a
$D^* \bar{D}$ molecule~\cite{Guo:2013xga}.
 
\subsection{Light quark notation}
\label{subsec:light-quark-notation}

Here we will use instead a more minimalistic notation,
in which we only take into account the light quark component
of the heavy hadron, i.e. the ``brown muck''.
The heavy quark fields within a heavy hadron act as a spectator,
where its major role is to provide a large effective mass
for the light quark attached to it.
The fact is that we can prescind of writing the heavy hadron superfields and
concentrate instead in the ``brown muck'', that is:
\begin{eqnarray}
  \mathcal{M} \rightarrow q_L \, ,
\end{eqnarray}
where $\mathcal{M}$ is the original superfield and $q_L$ is
a non-relativistic field containing
the light spin degrees of freedom.

We can now consider the contact-range Lagrangian without derivatives,
which is an illustrative example, in this notation.
For the $H_{\bar Q}$ heavy mesons and $T_{QQ}$ doubly heavy baryons we have
\begin{eqnarray}
  \mathcal{L} = C_a (q_L^{\dagger} q_L) (q_L^{\dagger} q_L) + C_b\,
  (q_L^{\dagger} \vec{\sigma}_L q_L)\, \cdot \,(q_L^{\dagger} \vec{\sigma}_L q_L)
  \, ,
  \label{eq:4L-contact}
\end{eqnarray}
where $\sigma_L$ refers to the spin of the light quark degrees of freedom.
From this Lagrangian we derive the following non-relativistic
contact-range potential
\begin{eqnarray}
  V = C_a + C_b \vec{\sigma}_{L1} \cdot \vec{\sigma}_{L2} \, .
  \label{eq:4L-VC}
\end{eqnarray}
Now we simply have to provide a series of rules for translating
the light quark spin operators into the spin operators for the heavy hadrons.
In the case of the heavy mesons the rules are
\begin{eqnarray}
  \langle P | \vec{\sigma}_L | P \rangle &=& 0 \, , \\
  \langle P | \vec{\sigma}_L | P^* \rangle &=& \vec{\epsilon} \, , \\
  \langle P^* | \vec{\sigma}_L | P^* \rangle &=& \vec{J} \, , 
\end{eqnarray}
where $\vec{\epsilon}$ is the polarization vector of the $P^*$ heavy meson
and $\vec{J}$ is the spin-1 matrix.
For the doubly heavy baryons we have~\footnote{For the charmed-bottom baryons,
  besides the $\Xi_{bc}$ and $\Xi_{bc}^*$ configurations in which the spin of
  the heavy diquark is $S_H=1$, there is also a $\Xi_{bc}'$ configuration
  in which the heavy diquark spin is $S_H=0$. For the $\Xi_{bc}'$ the rule
  is $\langle \Xi_{bc}' | \vec{\sigma}_L | \Xi_{bc}' \rangle = \vec{\sigma}$.}
\begin{eqnarray}
  \langle \Xi_{QQ} | \vec{\sigma}_L | \Xi_{QQ} \rangle &=&
  -\frac{1}{3}\,\vec{\sigma} \, , \\
  \langle \Xi_{QQ} | \vec{\sigma}_L | \Xi_{QQ}^* \rangle &=&
  -\frac{2}{\sqrt{3}}\,\vec{S} \, , \\
  \langle \Xi_{QQ}^* | \vec{\sigma}_L | \Xi^*_{QQ} \rangle &=&
  +\frac{2}{3}\,\vec{\Sigma} \, , 
\end{eqnarray}
where $\sigma$ are the Pauli matrices, which serve as the spin operators for
the $\Xi_{QQ}$ spin-$1/2$ doubly heavy hadrons, $\vec{S}$ is a matrix
for spin-$1/2$ to spin-$3/2$ transitions (the explicit form of which
can be consulted in Ref.~\cite{Lu:2017dvm}) and $\vec{\Sigma}$ are
the spin-$3/2$ matrices.
From these rules and the contact-range potential of Eq.~(\ref{eq:4L-VC})
it is easy to check that we arrive to the same potentials
that we have derived previously in a more laborious way
in Tables \ref{tab:1}, \ref{tab:2} and \ref{tab:3}.
The point is that we can write them more compactly simply as
$C_a + C_b \, \vec{\sigma}_{L1} \cdot \vec{\sigma}_{L2}$.
This is the advantage of the notation proposed here.

\section{A formal equivalence between nucleons and heavy mesons}
\label{sec:equivalence}

Nucleons are spin-$1/2$ non-relativistic fields. The contact-range
Lagrangian can be written as
\begin{eqnarray}
  \mathcal{L} =
  C_S (N^{\dagger} N)\,(N^{\dagger} N) + C_T
  (N^{\dagger} \vec{\sigma} N)\, \cdot \,(N^{\dagger} \vec{\sigma} N) \, ,
\end{eqnarray}
which is formally identical to the corresponding Lagrangian
for the S-wave heavy mesons and doubly heavy baryons,
see Eq.~(\ref{eq:4L-contact}), after the exchanges
\begin{eqnarray}
  C_a \leftrightarrow C_S \quad , \quad C_b \leftrightarrow C_T
  \quad , \quad \vec{\sigma}_L \leftrightarrow \vec{\sigma} \, ,
\end{eqnarray}
where for simplicity we have ignored isospin and
the statistics of the hadrons involved~\footnote{
  Actually, this can be easily taken into account by writing the contact-range
  Lagrangian with projectors, i.e. $\mathcal{L} = \sum_\alpha C_\alpha
  (N^T P_{\alpha} N)^{\dagger} (N^T P_{\alpha} N)$, with $\alpha$ the spin-isospin
  channel we are considering and $P_{\alpha}$ a suitable projector
  (plus a similar expression for the light-quark subfield).
  For the two-nucleon case, which are fermions, we end up with
  two S-wave coupling, while for the two heavy-hadron case,
  which can be either fermions ($\Xi_{QQ}$, $\Xi_{QQ}^*$)
  or bosons ($P$, $P^*$), we end up with four S-wave
  couplings.
}.
This is not surprising if we take into account that in both cases we have
a spin-$\frac{1}{2}$, isospin-$\frac{1}{2}$ non-relativistic field, where
the only formal difference is that nucleons belong to the $8$
representation of SU(3)-flavor, while the S-wave heavy
hadrons belong to the $3$ representation.
This last detail only manifest if we consider systems with strangeness.
Notice that the substitution rule for the couplings is expected to work
at the formal level, where the specific numerical values
can be different to each other.
We do not offer though a direct comparison of these values,
the reason being that they depend on the regularization scheme and
the cutoff chosen.
But we can advance that while for the two nucleon system the contact-range
couplings are relatively strong in both the spin-singlet and -triplet
configurations, phenomenological considerations suggest that
in the two doubly-charmed baryon system only the spin-triplet
configuration will be strongly attractive,
see Sect.~\ref{sec:charm-triplet}.

This equivalence is not limited to the contact-range interactions,
but extends to the pion interactions. If we consider vertices involving
one pion field and two hadron fields, we have
\begin{eqnarray}
  \mathcal{L}_{\pi NN} &=& \frac{g_A}{\sqrt{2} f_{\pi}}\,
  N^{\dagger} \vec{\sigma}_L \cdot \vec{\nabla}({\tau_a \pi_a})\,N \, ,
  \label{eq:LOPE-NN} \\
  \mathcal{L}_{\pi q_L q_L} &=& \frac{g_1}{\sqrt{2} f_{\pi}}\,
  q_L^{\dagger} \vec{\sigma}_L \cdot \vec{\nabla}({\tau_a \pi_a})\,q_L \, ,
  \label{eq:LOPE-qq}
\end{eqnarray}
with $a$ an isospin index and where we have used
the normalization $f_{\pi} \simeq 132\,{\rm MeV}$.
This implies that the one pion potential will be identical
in both systems modulo the new substitution rule
\begin{eqnarray}
  g_1 \leftrightarrow g_A \, , \label{eq:sub1}
\end{eqnarray}
where while $g_A = 1.26$, we have that $g_1 = 0.60$ (a value deduced from the
$D^* \to D \pi$ decay width~\cite{Ahmed:2001xc,Anastassov:2001cw}),
indicating that pion interactions are considerably weaker
in the heavy meson system when compared to nuclear
physics~\cite{Fleming:2007rp,Valderrama:2012jv}.
Now if we consider the Weinberg-Tomozawa terms (i.e. the leading terms
involving two pion fields and two hadron fields):
\begin{eqnarray}
  \mathcal{L}_{\pi \pi N N} &=& - \frac{1}{2 f_{\pi}^2}\,
  N^{\dagger} (\epsilon_{abc} \, \tau_a  {\pi}_b \,
  \partial_0 {\pi}_c) N \, , \\
  \mathcal{L}_{\pi \pi q_L q_L} &=& - \frac{1}{2 f_{\pi}^2}\,
  q_L^{\dagger} (\epsilon_{abc} \, \tau_a  {\pi}_b \,
  \partial_0 {\pi}_c) q_L \, ,
\end{eqnarray}
they happen to be identical (equivalently, we could have simply noticed
that the strength of the $NN\pi\pi$ and $DD\pi\pi$ vertices is the same).

To summarize, there is a formal equivalence.
Of course there is no actual equivalence because the couplings are different
in each case, the symmetry requirements can change and the light quarks
in the heavy meson case belong to the $3$ representation of
the SU(3)-flavor group, instead of the octet representation,
which is the case for the nucleons.
However as far as we are limited to pions and non-strange hadrons,
the equivalence holds.
The underlying reason for this equivalence is that nucleons and the light quarks
within heavy mesons and doubly-heavy baryons belong to the same irreducible
representations of the spin and isospin groups.

\section{Pion exchange in the light quark formalism}
\label{sec:pions}

Now we apply the light quark formalism to derive the potential between
(S-wave) heavy meson and the (S-wave) doubly heavy baryons.
For this we simply exploit the formal equivalence with nuclear physics
that we have explained in the previous section.
In fact no calculation is required (but a few caveats will be in order).

\subsection{One pion exchange}

The one pion exchange (OPE) potential for two nucleons is obtained from
the Lagrangian of Eq.(\ref{eq:LOPE-NN}), which leads to the well-known result
\begin{eqnarray}
  V_{\rm OPE}(\vec{q}) = - \frac{g_A^2}{2 f_{\pi}^2}\,
  \vec{\tau}_1 \cdot \vec{\tau}_2\,
  \frac{\sigma_{1} \cdot \vec{q} \, \sigma_{2} \cdot \vec{q}}{q^2 + m_{\pi}^2}
  \, .
\end{eqnarray}
From the substitution rules of Eq.(\ref{eq:sub1}),
the OPE between the two light quarks
within a heavy hadron can be directly written as
\begin{eqnarray}
  V_{\rm OPE}(\vec{q}) = - \frac{g_1^2}{2 f_{\pi}^2}\,
  \vec{\tau}_1 \cdot \vec{\tau}_2\,
  \frac{\sigma_{L1} \cdot \vec{q} \, \sigma_{L2} \cdot \vec{q}}{q^2 + m_{\pi}^2}
  \, .
\end{eqnarray}
If we use the rules for translating the light-quark spin operators into
the heavy-hadron spin operators, we will obtain the standard representation
of the OPE potential for heavy mesons~\cite{Valderrama:2012jv}.

\subsection{Leading two pion exchange}

\begin{figure*}
  \begin{center}
    \includegraphics[height=2.25cm]{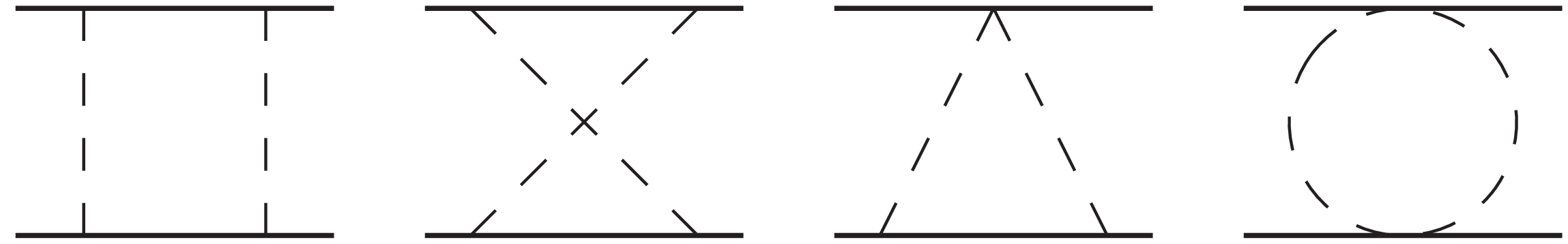}
    \end{center}
  \caption{Diagrams contributing to the leading TPE potential, which includes
    (from left to right) the planar box, crossed box, triangle and football
    diagrams. The solid and dashed lines represent the light-quark subfield
    and the pion field, respectively.
  }
\label{fig:1}       
\end{figure*}

The same idea applies to the leading two-pion exchange (TPE) potential
in nuclear EFT.
The diagrams contributing to the leading TPE potential are depicted
in Fig.~\ref{fig:1}, where we notice that there is no actual need
to recalculate these diagrams as we can directly employ
the results from nuclear EFT~\footnote{The Feynman rules are identical
  in both cases because the same formalism is being used
  for the heavy fields, independently of whether they are
  heavy hadrons~\cite{Isgur:1989vq,Isgur:1989ed}
  or nucleons~\cite{Jenkins:1990jv,Bernard:1992qa}.}.
For two-nucleons the leading TPE potential is written as
\begin{eqnarray}
  V_{\rm TPE-L}(\vec{q)} &=& W_C(\vec{q})\,\vec{\tau}_1 \cdot \vec{\tau}_2
  + V_S(\vec{q})\,\vec{\sigma}_1 \cdot \vec{\sigma}_2 \nonumber \\ &+&
  V_T(\vec{q})\,\vec{\sigma}_1 \cdot \vec{q} \, \vec{\sigma}_2 \cdot \vec{q} \,
  \, ,
\end{eqnarray}
where $W_C$, $V_S$ and $V_T$ are given by
\begin{eqnarray}
  W_C(\vec{q}) &=& -\frac{L(q)}{192 \pi^2 f_{\pi}^2}\,\Big[
    4 m_{\pi}^2\,\left( 5 g_A^4 - 4 g_A^2 - 1 \right) \nonumber \\
    && \quad + q^2 \,\left( 23 g_A^4 - 10 g_A^2 - 1 \right) \nonumber \\
    && \quad + \frac{48 g_A^4 m_{\pi}^2}{w(\vec{q})} \Big] \, , \\
  V_T(\vec{q}) &=& -\frac{1}{q^2}\,V_S(\vec{q}) =
  -\frac{3 g_A^3 L(q)}{32 \pi^2 f_{\pi}^2} \, ,
\end{eqnarray}
with $L(q)$ and $w(q)$ defined as
\begin{eqnarray}
  L(q) &=& \frac{w(q)}{q}\,\log{\left( \frac{w(q)+q}{2 m_{\pi}} \right)} \, , \\
  w(q) &=& \sqrt{4 m_{\pi}^2 + q^2} \,  ,
\end{eqnarray}
see for instance Ref.~\cite{Machleidt:2011zz} for an explicit derivation.
For the two pieces of ``brown muck'' (i.e. the two light quarks)
inside the heavy mesons and doubly heavy baryons
the leading TPE potential is identical after the substitutions
\begin{eqnarray}
  g_A \leftrightarrow g_1 \quad , \quad
  \vec{\sigma}_i \leftrightarrow \vec{\sigma}_{Li} \, ,
\end{eqnarray}
and that is all about it, mostly, except for a detail
that we will explain below.

\begin{figure}
  \begin{center}
    \includegraphics[height=2.25cm]{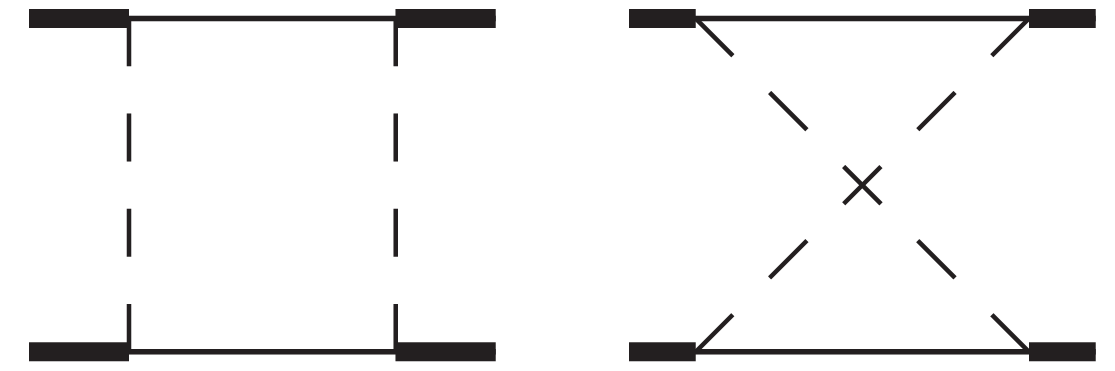}
    \end{center}
  \caption{
    The range of the planar-box, crossed box and triangle diagrams
    will be enhanced if the intermediate heavy hadron states
    are lighter than the initial and final states.
    The most illustrative example are the box diagrams: in the particular
    case of the $D^* D^*$ leading TPE potential,
    the intermediate $DD \pi \pi$ states are almost on the mass shell,
    which in turn leads to a large enhancement of
    the range of these diagrams.
    In the figure the thick and narrow solid lines represent the $D^*$ and $D$
    charmed mesons and the dashed line the pions.
  }
\label{fig:2}       
\end{figure}

The equivalence of the potentials is only true in the heavy quark limit,
for which the $P$ and $P^*$ heavy mesons and the $\Xi_{QQ}$
and $\Xi_{QQ}^*$ doubly heavy baryons are degenerate.
For a finite heavy quark mass, the mass of these heavy hadrons
will not be identical, where~\cite{Savage:1990di,Hu:2005gf}
\begin{eqnarray}
  m(P^*) - m(P) = \frac{4}{3}\,\left( m(\Xi_{QQ}^*) - m(\Xi_{QQ}) \right)
  = \Delta_Q \, , \label{eq:DeltaQ}
\end{eqnarray}
with the mass gap $\Delta_Q$ scaling as $\Lambda_{\rm QCD}/m_Q$.
The existence of this mass gap is mostly harmless if
\begin{eqnarray}
  \Delta_Q \ll m_{\pi} \, ,
\end{eqnarray}
in which case the mass gap will entail small corrections to
the leading TPE potential we have derived for $\Delta_Q = 0$.
But if this condition is not met, the mass gap will play an important role
in diagrams involving heavy hadrons and pion loops, which include
the triangle diagrams but most notably the planar and crossed
box diagrams.
For instance, if we consider the $D^*$ and $D$ mesons the box diagrams
where the initial and final state is the $D^*D^*$ system will
imply a $DD\pi\pi$ loop, as  depicted in Fig.~\ref{fig:2}.
It happens that the  $DD\pi\pi$ intermediate state  is roughly
at the same energy level as the initial and final $D^*D^*$
\begin{eqnarray}
  m(D^* D^*) \simeq m(DD\pi\pi) \, , 
\end{eqnarray}
and this implies that the range of the box diagrams will be incredibly
enhanced in the $D^* D^*$ system.
This effect will be however more suppressed in other two heavy hadron systems
owing to the smaller energy gaps.
For instance, the energy gap for the doubly charmed baryons
is expected to be about $3/4$ of that of the charmed mesons,
see Eq.~(\ref{eq:DeltaQ}). Conversely, in the bottom sector
the energy gaps are about $1/3$ of those
in the charmed sector.
Finally it is worth mentioning the existence of previous calculations
of the leading TPE potential explicitly taking into account
the mass gap for the the $D D^{*}$ and $B^* B^*$
systems~\cite{Xu:2017tsr,Wang:2018atz}.

\section{Does a charming nuclear landscape emerge?}
\label{sec:exotic}

The formal equivalence between the nucleon and the heavy hadron sectors
begs the question: are there exotic equivalents of
the standard nuclear landscape?
That is, are there nuclear landscapes composed of $\Xi_{cc}$,
$\Xi_{bc}$ or $\Xi_{bb}$ doubly heavy baryons instead of nucleons?

In this manuscript we will only consider the first possibility. i.e.
the hypothetical nuclear landscape composed of doubly charmed baryons.
For systems of $\Xi_{cc}$ baryons the answer is that probably there is no
equivalent of the nuclear landscape, but this is not completely settled:
the two- and three-body $\Xi_{cc}$ systems are in the limit between
binding and not binding and will deserve further
investigation in the future.
But as the number $A$ of $\Xi_{cc}$ baryons increases a big difference with
the nucleons manifest: the $\Xi_{cc}$ baryons are always electrically
charged, where the two isospin states correspond to
the $\Xi_{cc}^{+}$ and $\Xi_{cc}^{++}$ baryons.
As a consequence the Coulomb repulsion in a theoretical charming nuclear
landscape will increase much faster than in the standard nuclear
landscape. The likelihood of systems of $A$ doubly charmed
baryons binding will decrease quickly as $A$ increases.
This limitation is not present for the charm-beauty $\Xi_{bc}$ and
doubly beautiful $\Xi_{bb}$ baryons, yet we will not study
these two potential nuclear landscapes in this contribution.

The starting point is whether we have the same building blocks
as in nuclear physics.
Even though there is no necessarily a universal consensus,
here we propose the following basic blocks:
(i) nucleon-like particles (fermions with four internal states),
(ii) SU(4) Wigner symmetry and
(iii) a closeness to the unitary limit in the S-wave Wigner multiplets.
The choice of these three blocks is indeed very much influenced
by the idea of nuclear physics in the unitary limit
proposed in Ref.~\cite{Konig:2016utl}.

Doubly heavy baryons provide the first basic block:
the ground state $\Xi_{cc}$ baryon acts like a nucleon,
being a fermion and having the same spin and isospin quantum numbers.
The second and third blocks are however not necessarily
present for systems composed of $\Xi_{cc}$ baryons.
For determining whether this is the case we will rely on a phenomenological
model of hadron-hadron interactions, for which we choose the OBE model.
As we will see, for the $\Xi_{cc} \Xi_{cc}$ two-body system the singlet and
triplet scattering lengths are far from being similar, where only
the triplet is close to the unitary limit.

\subsection{A charming triplet without a charming singlet}
\label{sec:charm-triplet}

For determining the scattering lengths of the $\Xi_{cc} \Xi_{cc}$ two-body system
we will use the OBE model.
The OBE model is a physically compelling and phenomenologically successful
picture of the nuclear forces~\cite{Machleidt:1987hj,Machleidt:1989tm},
where the nuclear potential is generated by the exchange of a
few light mesons (including the pion, the $\sigma$,
the $\rho$ and the $\omega$).
Of course the determination of the scattering lengths in the OBE model
is phenomenological, i.e. it is subjected to unknown systematic
uncertainties that we do not know how to estimate.
There have been numerous applications of the OBE model
to hadronic molecules~\cite{Liu:2008tn,Sun:2011uh,Chen:2015loa,Liu:2018bkx,Wang:2019nwt},
where the particular OBE model we follow is the one developed
for heavy meson-meson and heavy meson-antimeson systems
in Ref.~\cite{Liu:2019stu}, which I have adapted here
for the doubly heavy hadrons.
The OBE model requires regularization, which is usually implemented by the
inclusion of a suitable form factor and a cutoff.
The version of the OBE model presented in Ref.~\cite{Liu:2019stu} determines
the value of the cutoff from the condition of reproducing
the mass of the $X(3872)$.
This idea is inspired by the renormalized OBE model
of Ref.~\cite{Cordon:2009pj}.
The explicit comparison between the calculations in the partially renormalized
OBE model of Ref.~\cite{Liu:2019stu} and previous EFT calculations
for hadron molecules (in particular the calculations of
Ref.~\cite{Guo:2013sya}) lead to similar results.
There is a previous study of the $\Xi_{cc} \Xi_{cc}$ system in the OBE
model~\cite{Meng:2017fwb}, but in it the cutoff is left to float
within a sensible range, which means that predictions
are less specific.

Concrete calculations for the doubly charmed baryons in the OBE model
of Ref.~\cite{Liu:2019stu} lead to
\begin{eqnarray}
  a_s = -1.1^{+0.5}_{-0.2}\,{\rm fm} \quad \mbox{and} \quad
  a_t = +14.2_{-9.7}^{+\infty (-7.5)}\,{\rm fm} \, . \nonumber \label{eq:a2} \\
\end{eqnarray}
The errors stem from assuming a $30\%$ uncertainty in the OBE potential,
see Ref.~\cite{Liu:2019stu} for details; the ${+\infty}$ ${(-7.5)}$
superscript indicates that the triplet scattering length
can change sign, with the corresponding value
in parenthesis.
It happens that the singlet is not particularly close to the unitary limit:
it is attractive, but only in moderation, leading to
a natural scattering length.
The triplet is however close to the unitary limit: though its exact fate is
difficult to predict, the unitary limit is within the uncertainties of the
phenomenological model we are using.
That is, the charming deuteron is as likely to be bound as unbound.
The situation is however not completely analogous to standard nuclear physics,
where we are close to the unitary limit in both the singlet
and the triplet channels.
If the previous calculations are on the right track, charming nuclear physics
lie in what we can tentatively call the {\it semi-unitary limit}, i.e.
the limit in which only the singlet or the triplet displays
a unnaturally large scattering length.

\subsection{The uncertain fate of the charming triton}

Systems of doubly heavy baryons do not seem to display Wigner-SU(4) symmetry.
This is definitely inconvenient when dealing with the $A = 3,4$ cases,
with $A$ the number of $\Xi_{cc}$ baryons.
The reason is that with Wigner-SU(4) symmetry the $A = 3,4$ systems reduces
to systems of identical bosons~\cite{Bedaque:1998km,Bedaque:1999ve,Konig:2016utl}.
Without Wigner-SU(4) the calculations are more involved.

While for the initial exploration of the $A=2$ systems
we have used a phenomenological model (the OBE model),
for the $A \geq 3$ systems it is more convenient
to use the pionful EFT formalism instead.
The pionful EFT for the doubly heavy baryons is however different
in two important respects to the pionful EFT for nucleons.
The first difference is the role of pions: pion couplings are particularly
weak for the $\Xi_{cc}$ baryons.
By particularizing the light-quark Lagrangian of Eq.~(\ref{eq:LOPE-qq}) to
the doubly heavy $\Xi_{QQ}$ baryons we obtain
\begin{eqnarray}
  \mathcal{L}_{\pi \Xi_{QQ} \Xi_{QQ}} &=& -\frac{1}{3}\frac{g_1}{\sqrt{2} f_{\pi}}\,
  \Xi_{QQ}^{\dagger} \vec{\sigma}
  \cdot \vec{\nabla}({\tau_a \pi_a})\,\Xi_{QQ}\, ,
  \label{eq:LOPE-XiQQ}
\end{eqnarray}
which implies that the effective axial coupling of the pions to the doubly
heavy baryons is $g_1' = - g_1 / 3$ ($\simeq -0.2$).
The relative strength of the OPE potential depends on the square of this
axial coupling, ${(g_1')}^2$, and the reduced mass of the system.
If we take into account that the OPE potential is perturbative
in the charmed meson-antimeson system~\cite{Valderrama:2012jv},
the previous two factors indicate this conclusion to be even more
applicable to systems of two doubly charmed baryons~\footnote{
  The reduced mass of a system of two $\Xi_{cc}$ baryons is about twice
  the one of a system of two charmed mesons. The square of the
  axial coupling is nine times smaller. Compounded together,
  the conclusion is that OPE is $4-5$ times weaker
  in the $\Xi_{cc} \Xi_{cc}$ system
  than in the $D^* D^*$ one.}.
As a consequence, pions are subleading: in the pionful EFT for
the $\Xi_{cc}$ baryons pions are perturbative, which means that
these systems follow a variation of the Kaplan, Savage and
Wise power counting~\cite{Kaplan:1998tg,Kaplan:1998we}.
The pionful EFT at leading order (LO) involves contact interactions only and
is thus identical to the LO pionless EFT,
except for the larger radius of convergence.
The other difference is the singlet and triplet couplings:
only the triplet scattering length is large
in comparison to the range of the pion.
Thus at LO in principle it is enough to only include
the triplet scattering length in the EFT calculation,
with the singlet scattering length entering
at subleading orders.

The starting point are the Faddeev equations for the three doubly heavy baryon
system, which are of course identical to the ones of the three nucleon system.
Besides in pionless EFT there are only contact-range interactions.
This usually translates into separable interactions if the contact-range
interactions are regularized with a suitable regulator,
for instance a Gaussian regulator
\begin{eqnarray}
  \langle q' | V_C | q \rangle &=& C(\Lambda)
  g(q') g(q) \nonumber \\
  &=& C(\Lambda)
  \,e^{-(q'^{2n} / \Lambda^{2n})}\,e^{-(q^{2n} / \Lambda^{2n})} \, ,
\end{eqnarray}
with $g(k)$ a regulator, which in the second line we take to be a Gaussian
$g(k) = e^{-(k/\Lambda)^{2n}}$, where $\Lambda$ is the cutoff.
For separable potentials the T-matrix is also separable and reads
\begin{eqnarray}
  \langle q' | T(Z) | q \rangle &=& \tau(Z)\,g(q') g(q) \, ,
  \label{eq:T-separable}
\end{eqnarray}
where $Z$ refers to the energy.
As is well known, the Faddeev equations also take a particularly simple form
for separable interactions, which in the case of the three nucleon
system can be consulted for instance in Ref.~\cite{Meier:1984hj}.
For completeness we briefly review the resulting Faddeev equations,
where the specific notation we use has been influenced
by Ref.~\cite{Garcilazo:1984rx}.
We begin with the Faddeev decomposition of the wave function
\begin{eqnarray}
  \Psi_{3B} = \sum_{(ijk)} && \Big[
  \psi_s(p_k, q_{ij}) | {0}_{ij} \otimes \frac{1}{2}_k \rangle_S
  | {1}_{ij} \otimes \frac{1}{2}_k \rangle_T
  \nonumber \\
  && + \psi_t(p_k, q_{ij}) | {1}_{ij} \otimes \frac{1}{2}_k \rangle_S
  | {0}_{ij} \otimes \frac{1}{2}_k \rangle_T \Big]
  \, , \nonumber \\
\end{eqnarray}
where $\psi_{s(t)}$ are the singlet and triplet Faddeev components,
$p_k$ and $q_{ij}$ are the Jacobi momenta,
$(ijk)$ is an even permutation of $123$,
$| S_{ij} \otimes \frac{1}{2} \rangle_{S}$ refers to the spin wave functions
(with particles $ij$ coupling to spin $S_{ij}$) and
the isospin wave function is defined analogously but uses the subscript ``T''.
Notice that we are not explicitly indicating the total
spin/isospin of the three body system,
which we simply set to be $1/2$.
For a separable potential the Faddeev components take the simple form
\begin{eqnarray}
  \psi_{s(t)}(p, q) = \frac{f_{s(t)}(p) g(q)}
      {Z - \frac{\vec{k}_1^2}{2 m_1} -
        \frac{\vec{k}_2^2}{2 m_2} - \frac{\vec{k}_3^2}{2 m_3}
      } \, , 
\end{eqnarray}
with $\vec{k}_i$, $m_i$ the momenta and the masses of particles $i = 1,2,3$,
where the momenta fulfill the condition
$\vec{k}_1 + \vec{k}_2 + \vec{k}_3 = 0$.
In the isospin symmetric limit we take $m_i = m(N)$ or $m_i = m(\Xi_{QQ})$
depending on whether we are considering nucleons or
doubly heavy baryons.
The $f_{s}$ and $f_t$ components of the wave function follow
the coupled-channel, reduced Faddeev equations
\begin{eqnarray}
  f_s(p_1) &=& 2\tau_s(Z_{23})\,\int \frac{d^3 \vec{p}_2}{(2\pi)^3}\,
  B_{12}\,
  \left[ +\frac{1}{4} f_s(p_2) - \frac{3}{4} f_t(p_2) \right] \, , \nonumber \\
  \\
  f_t(p_1) &=& 2\tau_t(Z_{23})\,\int \frac{d^3 \vec{p}_2}{(2\pi)^3}\,B_{12}
  \left[ -\frac{3}{4} f_s(p_2) + \frac{1}{4} f_t(p_2) \right] \, ,
  \nonumber \\
\end{eqnarray}
where $\tau_s$ and $\tau_t$ are the energy-dependent components of
the separable T-matrix as defined in Eq.~(\ref{eq:T-separable}),
$Z_{23} = Z - \frac{p_1^2}{2 \mu_1}$
with $\mu_1 = m_1 (m_2 + m_3) / (m_1 + m_2 + m_3)$ and 
the function $B_{12}$ is written as
\begin{eqnarray}
  B_{12}(\vec{p}_1, \vec{p}_2) = \frac{g(q_1) g(q_2)}{Z -
    \frac{p_1^2}{2 m_1} - \frac{p_2^2}{2 m_2} - \frac{p_3^2}{2 m_3} 
  } \, ,
\end{eqnarray}
with $\vec{p}_3 = -(\vec{p}_1 + \vec{p}_2)$ and the $q_k$ defined as follows
\begin{eqnarray}
  \vec{q}_k = \frac{m_i \vec{p}_j - m_j \vec{p}_k}{m_i + m_j} \, ,
\end{eqnarray}
with $(ijk)$ an even permutation of $(123)$.
The previous set of homogeneous integral equations can be easily solved by
discretization methods, from which they are reduced
into an eigenvalue problem.
The value of $Z$ for which the eigenvalue is $1$ corresponds to
the bound state energy, while the associated eigenvector
can be interpreted as the bound state wave function.

The ${\rm LO}$ calculation of the charming triton now depends
on which contact-range interactions we include at ${\rm LO}$.
If we naively assume the singlet and triplet contact-range couplings
to enter the ${\rm LO}$ calculation, we find that for the central
values of the scattering length the charming triton is bound by
\begin{eqnarray}
  B_3 = (0.4 - 3.4)\,{\rm MeV} \, , \label{eq:B3}
\end{eqnarray}
where the range is the result of choosing the cutoff window
$\Lambda = 0.5-1.0\,{\rm GeV}$, with harder cutoffs
leading to more binding, and with the Gaussian
exponent set to $n=1$.
However the previous pionful EFT is not particularly good: the singlet
scattering length is more natural than unnatural.
Besides, if we include both scattering lengths the calculation of
the binding energy is actually not renormalizable because
this type of three body systems will suffer
from Thomas collapse~\cite{Thomas:1935zz}.

The most conservative pionful EFT for the $\Xi_{cc}$ baryons will be one
in which the singlet coupling is subleading and the triplet coupling
is leading.
In this pionful EFT the $\rm LO$ Faddeev equations only involve
the triplet channel and it happens that for the central value of
the triplet scattering length we have obtained,
i.e. $a_t = 14.2\,{\rm fm}$,
the charming triton is in the limit between binding and not binding.
In particular it can be shown that the charming triton binds
for triplet scattering lengths fulfilling the condition
\begin{eqnarray}
  a_t \leq (11.9-14.5)\,{\rm fm} \, ,
\end{eqnarray}
for the cutoff window $\Lambda = 0.5-1.0\,{\rm GeV}$.
That is, the charming triton might bind at LO.

Finally, there is a very important detail to consider: Coulomb repulsion.
The particle content of the charming triton is
$\Xi_{cc}^{+} \Xi_{cc}^{+} \Xi_{cc}^{++}$,
in which Coulomb repulsion acts on every possible particle pair.
This factor points against the formation of the charming triton.
If the charming triton had a similar spatial configuration to that of
the standard triton and the $^3{\rm He}$ nucleus, we should expect $5$
times more repulsion than for $^3{\rm He}$ (that is $3.5-5.0\,{\rm MeV}$
against $0.7-1.0\,{\rm MeV}$ for $^3{\rm He}$, as extracted
from nuclear EFT~\cite{Konig:2015aka}).
The size of these three-body systems can be approximated by
$R_3 \sim 1/\sqrt{2 \mu_1 B_3 }$, which yields $(1.5-4.5)\,{\rm fm}$
and $1.9\,{\rm fm}$ for the charming and standard triton,
respectively~\footnote{Here it is interesting to notice
  that the $\Xi_{cc}^{++}$ is about half the size of the proton:
  while the electromagnetic radius of the later is about
  $\sqrt{{\langle r^2_{e.m.} \rangle}_p} \simeq 0.83\,{\rm fm}$~\cite{Xiong:2019umf},
  for the former we have $0.4\,{\rm fm}$ according to
  the lattice calculation of Ref.~\cite{Can:2013tna}.
  This implies that the charming triton is closer to the universal limit
  than the standard triton, as its size is comparatively larger
  in relation to its constituents.
}.
This suggests that the charming triton might very well be on the brink
between binding and not binding, with its actual fate depending
on the interplay among its size, its spatial configuration
and the repulsive contribution from the Coulomb potential.

\subsection{What about the charming alpha particle?}

There is the possibility that the charming nuclear landscape is Brunnian,
i.e. the three body system is unbound, but the four body system binds,
see Ref.~\cite{Kirscher:2017xpj} for a brief and clear exposition.
This is not necessarily unlikely, particularly if we consider the differences
in the $A = 3$ and $A = 4$ systems and then compare them
with standard nuclear physics.

The argument is as follows.
First, for comparison purposes, we will consider a system of $A$ identical
bosons interacting via short-range forces.
The Schr\"odinger equation can be written as
\begin{eqnarray}
  \left[ H_0 + \frac{A (A-1)}{2} \bar{V} \right] \, \Psi_A = E \, \Psi_A \, ,
  \label{eq:A-bosons}
\end{eqnarray}
where $\bar{V}$ represents the average of the potential
for all the possible interacting pairs.
For $A=3$ and resonant two-body interactions this system displays a
characteristic discrete geometric scaling known as the Efimov
effect~\cite{Efimov:1970zz}, in which there is a tower of
bound states for which the ratio of the binding energies
of the $n$-th and $(n+1)$-th excited state is $E_n / E_{n+1} = (22.7)^2$.
Calculations have shown that this type of discrete spectrum
persist for $A = 4$, $5$ and $6$~\cite{Hammer:2006ct,vonStecher:2011zz,Gattobigio:2011ey} and, though we do not know this for sure,
this might very well extend to larger $A$.
There is a coordinate space explanation of the Efimov effect by
Fedorov and Jensen~\cite{PhysRevLett.71.4103} that
we will present here in a simplified form:
for $A=3$ there are 3 Faddeev components and 3 instances of the short-range
potential $V$, which in the zero-range limit generates a boundary condition
for each Faddeev component which in turn translates
into the discrete scaling we know.
This explanation can be extended to $A \geq 4$: for $A = 4$ there
are 6 Faddeev components, which can be further subdivided
into 12 ``K-type'' and 6 ``H-type'' Faddeev-Yakubovsky
components (that correspond to the different asymptotics
of the four-body system, see Ref.~\cite{Ciesielski:1998sy}
for a clear exposition).
The 6 instances of the short-range potential $V$ generate
in the zero-range limit a boundary condition for each of
the Faddeev components of $\Psi_A$, which are formally identical
to the boundary conditions in the $A = 3$ case
and lead to the same scaling factor.
For $A \geq 5$ the argument will follow the same lines: there are $A(A-1)/2$
instances of the short-range potential and $A(A-1)/2$ Faddeev
components. Each instance of the potential generates
a boundary condition for each one of the Faddeev component, and
we might end up with the same scaling as in the $3$ and $4$ body cases,
though in the absence of concrete calculations this is merely a conjecture.

The same argument applies for a system of $(A-1)$ identical bosons plus
one non-identical particle, where the bosons only interact
with the non-identical particle but not among themselves.
In this case we write the Schr\"odinger equation as
\begin{eqnarray}
  \left[ H_0 + (A-1) \bar{V} \right] \, \Psi_A = E \Psi_A \, ,
  \label{eq:A-heteronuclear}
\end{eqnarray}
where $\bar{V}$ represents the average of the potential
for all the possible interacting pairs, i.e. the interaction
between the non-identical particle and each of the $(A-1)$ bosons.
For $A = 3$ this system also displays the Efimov effect,
provided that the interactions are close to the unitary limit.
If the non-identical particle has the same mass as the two bosons,
the scaling factor will be considerably larger than in the
three boson system, where the ratio of the binding energies of
two consecutive states is $E_n / E_{n+1} = (1986.1)^2$.
This ratio will reduce (increase) if the non-identical particle
is lighter (heavier) than the bosons.
The same boundary condition argument as in the $A \geq 4$ boson system
might apply, but with the number of Faddeev components reduced to $(A-1)$.
This suggest that the discrete scaling is likely to extend for $A \geq 4$,
but this has only been explicitly checked for $A=4$ heteronuclear
systems with a large mass imbalance~\cite{PhysRevLett.108.073201,PhysRevLett.113.213201}.

For the standard alpha particle, the $^4{\rm He}$ nucleus, if we consider the
pure S-wave piece of the wave function it happens that the spin and
isospin wave function is totally antisymmetrical.
If we are in pionless EFT and write the Schr\"odinger equation for this
totally antisymmetric spin and isospin wave function
we obtain the following
\begin{eqnarray}
  [ H_0 + 3 \bar{V}_s + 3 \bar{V}_t ] \Psi_A = E \Psi_A \, ,
\end{eqnarray}
which in the Wigner SU(4) limit becomes identical to the corresponding
Schr\"odinger equation for the four boson system, Eq.~(\ref{eq:A-bosons}).
If we later apply the unitary limit then it becomes Efimov-like
with a scaling factor of $22.7$.
For the charming alpha particle, the conjectured $4 \Xi_{cc}$ bound state,
he Schr\"odinger equation reads
\begin{eqnarray}
  [ H_0 + 3 \bar{V}_t ] \Psi_A = E \Psi_A \, ,
\end{eqnarray}
which indicates half the total attraction as in the $^4{\rm He}$ system
(notice that we have ignored the attraction in the singlet channel
because it is expected to be considerably weaker
than in the triplet).
This is strikingly similar to the Schr\"odinger equation of
the heteronuclear system of Eq.~(\ref{eq:A-heteronuclear}),
which in the unitary limit is Efimov-like
with a scaling factor of $1986.1$.
Though the charming alpha particle is not completely analogous to
the $(A-1)$ bosons plus one particle system, it is nonetheless
similar enough as to conjecture that in the absence of
long-range Coulomb repulsion it might also be Efimov-like
with the aforementioned $1986.1$ discrete scaling
factor~\footnote{Notice that we are not making the explicit distinction
  between the three- and four-body Efimov effect that is sometimes
  done in the literature.}.
This conjecture is to be checked with concrete calculations
(requiring fantastically large scattering lengths or
cutoffs~\footnote{Preliminary numerical explorations are being conducted
  by J. Kirscher, S. K\"onig and C.-J. Yang, though no definite
  conclusion has been reached yet.
}).
The scaling factor is too large and even if the $A=4$ system was Efimov-like,
no Efimov state could be ever realistically expected to be observed,
particularly once we take into account Coulomb effects.
Yet the importance of the Efimov effect is a different one: its presence
will signal the possibility of Thomas collapse, which in turn will imply
more attraction than expected for the charming alpha particle.
Whether this additional attraction will compensate for the relatively large
Coulomb repulsion is to be seen: the charming alpha particle is a
$\Xi_{cc}^+ \Xi_{cc}^+ \Xi_{cc}^{++} \Xi_{cc}^{++}$ bound state,
from which we expect $13$ times more repulsion than
in the standard alpha particle if we assume
the same spatial configuration.
The semi-unitary limit might also has interesting ramifications
from the point of view of its EFT description:
if this type of system does indeed display discrete scaling,
it will require either a three- or four-body force for its renormalization.
But while the standard unitary limit is equivalent to the three-
and four-boson systems, where the three-body force enters
at LO~\cite{Bedaque:1998km,Bedaque:1999ve} and
the four-body force at next-to-leading order~\cite{Bazak:2018qnu},
the semi-unitary limit will probably exhibit
a more involved power counting.
It is nonetheless an interesting problem to look at.

\subsection{Heavy-quark fusion}

At first sight we might consider that the $A$-body bound states of
$\Xi_{cc}$ baryons are stable with respect to the strong force
(the $\Xi_{cc}$ decays weakly), but this is not the case.
Recently, Karliner and Rosner~\cite{Karliner:2017elp} proposed
the idea of a heavy-quark analogue of nuclear fusion.
The example they considered is the $\Lambda_c \Lambda_c$ system, in which
the two charmed quarks can in principle combine for the system
to decay into $\Xi_{cc} N$, with $N$ a nucleon.
The same decay can in principle happen for $\Xi_{cc}$ bound states.
For instance, the charming deuteron can undergo heavy-quark fusion as follows:
\begin{eqnarray}
  \Xi_{cc} \Xi_{cc} \to \Omega_{ccc} \Lambda_c \, ,
\end{eqnarray}
where from the experimental masses of the $\Xi_{cc}$ and $\Lambda_c$
baryons~\cite{Tanabashi:2018oca}
and the lattice QCD mass calculation of the $\Omega_{ccc}$
baryon~\cite{Brown:2014ena},
we expect this reaction to release $Q \sim 160\,{\rm MeV}$.
This shows that the exotic nuclear landscapes are actually not
stable under the strong interaction.
Of course the question is how important is this type of decay, but taking
into account that this is mediated by a short-range operator
(where the range is probably shorter than the size of
the doubly charmed baryons), while the $A=2$, $3$, $4$ charming nuclei
we have discussed here are probably very weakly bound,
the expectation is that heavy-quark fusion will be a relatively
slow process generating a narrow width.

\section{Conclusions}
\label{sec:summary}

This manuscript proposes and exploits the idea that heavy hadron-hadron
interactions are formally equivalent to nuclear physics, particularly
in the case of S-wave heavy mesons and doubly heavy baryons.
This idea has been in the air since heavy hadron molecules were
initially conjectured on the basis of an analogy to the deuteron
and the nuclear forces that binds it~\cite{Voloshin:1976ap}.
Heavy hadrons, provided they contain light quarks, can exchange
light mesons ($\pi$, $\sigma$, $\rho$, $\omega$), which in turn
generate a potential that might be able to bind
these hadrons together.
This picture has inspired many of the subsequent theoretical investigations
about hadronic molecules~\cite{Manohar:1992nd,Tornqvist:1993ng,Ericson:1993wy}.
Here we have simply investigated this idea further, where by using a suitable
notation the previous analogy can be shown to be a formal equivalence.

This equivalence can be exploited to make calculations easier
or to generate theoretical predictions.
A first example is the calculation of the one- and two-pion exchange
potentials for heavy hadron-hadron system.
A second example is the possible existence of exotic nuclear landscape,
i.e. equivalents of the standard nuclei that are composed of
doubly heavy baryons instead of nucleons.
In particular, by making use of both phenomenological and EFT methods,
we have explored in more detail the nuclear landscape
generated by the $\Xi_{cc}$ doubly charmed mesons.
This nuclear landscape is unlikely to spawn beyond $A > 4$, owing to the
rapidly increasing Coulomb repulsion, even if it only comprises
$A = 2$, $3$, $4$ nuclei it will still be really interesting
from the theoretical point of view.
The charming nuclear landscape might be in the {\it semi-unitary limit},
in which only one of the two-body S-wave configurations (the singlet
or the triplet) is close to the unitary limit.
In contrast, standard nuclear physics might be understood
as an expansion around the unitary limit~\cite{Konig:2016utl},
where both the singlet and the triplet scattering lengths
are larger than the other characteristic scales of
the two-nucleon system.
The exotic nuclear landscapes composed of the $\Xi_{bc}$ and $\Xi_{bb}$
doubly-heavy baryons are likely to extend beyond $A = 4$,
but we have not considered them explicitly.

\section*{Acknowledgments}
I would like to thank Johannes Kirscher, Sebastian K\"onig,
Ubirajara van Kolck and Jerry Yang for discussions
concerning the semi-unitary limit.
This work is partly supported by the National Natural Science Foundation of
China under Grant No. 11735003, the fundamental Research Funds
for the Central Universities, and the Thousand Talents Plan
for Young Professionals.


%

%
%
%
%

\end{document}